\documentclass[prx,aps, twocolumn, amsmath,amssymb,superscriptaddress,longbibliography]{revtex4-1}

\usepackage{graphicx}
\usepackage{dcolumn}
\usepackage{bm}
\usepackage[sort&compress]{natbib}
\usepackage{xspace}
\usepackage{amsmath}
\usepackage{epstopdf}
\usepackage{amssymb}
\usepackage{epsfig}
\usepackage{color}

\begin{document}

\title{Acousto-optic and opto-acoustic modulation in piezo-optomechanical circuits}
\author{Krishna C. Balram}\email{krishna.coimbatorebalram@nist.gov}
\affiliation{Center for Nanoscale Science and Technology, National
Institute of Standards and Technology, Gaithersburg, MD 20899,
USA}\affiliation{Maryland NanoCenter, University of Maryland,
College Park, MD 20742, USA}
\author{Marcelo I. Davan\c co}
\affiliation{Center for Nanoscale Science and Technology, National
Institute of Standards and Technology, Gaithersburg, MD 20899,
USA}
\author{B. Robert Ilic}
\affiliation{Center for Nanoscale Science and Technology, National
Institute of Standards and Technology, Gaithersburg, MD 20899,
USA}
\author{Ji-Hoon Kyhm}
\affiliation{Quantum-functional Semiconductor Research Center, Dongguk University, Seoul 04620, South Korea }
\author{Jin Dong Song}
\affiliation{Center for Opto-Electronic Materials and Devices Research, Korea Institute of Science and Technology, Seoul 136-791, South Korea }
\author{Kartik Srinivasan} \email{kartik.srinivasan@nist.gov}
\affiliation{Center for Nanoscale Science and Technology, National
Institute of Standards and Technology, Gaithersburg, MD 20899, USA}

\date \today

\begin{abstract}
{Acoustic wave devices provide a promising chip-scale platform for efficiently coupling radio frequency (RF) and optical fields.  Here, we use an integrated piezo-optomechanical circuit platform that exploits both the piezoelectric and photoelastic coupling mechanisms to link 2.4~GHz RF waves to 194~THz (1550~nm) optical waves, through coupling to propagating and localized 2.4~GHz acoustic waves.  We demonstrate acousto-optic modulation, resonant in both the optical and mechanical domains, in which waveforms encoded on the RF carrier are mapped to the optical field. We also show opto-acoustic modulation, in which the application of optical pulses gates the transmission of propagating acoustic waves.  The time-domain characteristics of this system under both pulsed RF and pulsed optical excitation are considered in the context of the different physical pathways involved in driving the acoustic waves, and modelled through the coupled mode equations of cavity optomechanics.}
\end{abstract}

\maketitle

\setcounter{figure}{0}
\makeatletter
\renewcommand{\thefigure}{\@arabic\c@figure}

\setcounter{equation}{0}
\makeatletter
\renewcommand{\theequation}{\@arabic\c@equation}

\section{Introduction}
Nanomechanical systems have been widely used as transducers~\cite{cleland2013foundations} in sensing applications where an external force is mapped onto the motion of a mechanical resonator, which in turn is read out through electrical or optical means.  Coupling such nanomechanical systems to electromagnetic waves~\cite{aspelmeyer2014cavity} has further expanded the available experimental toolkit, enabling measurement imprecision below the standard quantum limit~\cite{teufel2009nanomechanical,schliesser2009resolved,wilson2015measurement}, sideband cooling~\cite{chan2011laser,teufel2011sideband} and parametric amplification~\cite{rokhsari2005radiation} of mechanical motion, and use of the mechanical resonator as an effective nonlinear medium for frequency conversion~\cite{hill2012coherent,dong2012optomechanical,liu2013electromagnetically}.  This latter role can be extended to situations in which radio frequency (RF) and optical waves are parametrically coupled to the same mechanical system. Achieving an RF-to-optical link is of relevance to quantum information science as a bridge between superconducting circuits whose qubits and resonators operate in the few GHz range~\cite{barends2014superconducting} and low propagation loss fiber-optical links in the 200~THz telecommunications band~\cite{andrews2014bidirectional,kurizki2015quantum}. It is also of significant interest for classical signal processing, where the low speed of sound and accompanying reduction in wavelength in comparison to an electromagnetic wave of the same frequency (a factor of $\approx10^{-5}$ in GaAs) allows one to create compact integrated filters, delay lines, and signal buffers~\cite{hashimoto2000surface}, while an optical interface enables encoding RF signals onto an optical carrier for long-distance information transmission~\cite{urick2015fundamentals}.

Piezoelectric optomechanical devices~\cite{bochmann2013nanomechanical,fong2014microwave,tadesse2014sub,li2015nanophotonic,balram2016coherent} provide one route for coupling RF and optical waves.  Our approach to such devices was recently described in Ref.~\onlinecite{balram2016coherent}, where we used phononic crystal waveguides to link the propagating surface acoustic waves (SAWs) generated by an RF interdigitated transducer (IDT) to an optomechanical cavity.  Here, we demonstrate the dynamic manipulation of optical and acoustic waves in this system.  In particular, we show acousto-optic modulation, in which waveforms encoded on an RF carrier are mapped to the optical domain, and opto-acoustic modulation, in which optical pulses gate the transmission of propagating acoustic waves.  This ability to dynamically control acoustic wave propagation through the application of optical fields is a novel functionality of these piezo-optomechanical systems, and may be particularly relevant considering the limited alternatives for achieving such functionality. Along with indicating the potential of these devices in optical and acoustic wave based signal processing, our measurements provide insights into how the different mechanisms by which we drive the optomechanical resonator (optically through photoelastic coupling or electrically through piezoelectric coupling to propagating acoustic waves) differ in their time-domain response.  This behavior, which includes signatures of dynamic back-action under optical excitation, is well-modelled by the coupled cavity optomechanical interaction equations.

\begin{figure*}
\includegraphics[width=\linewidth]{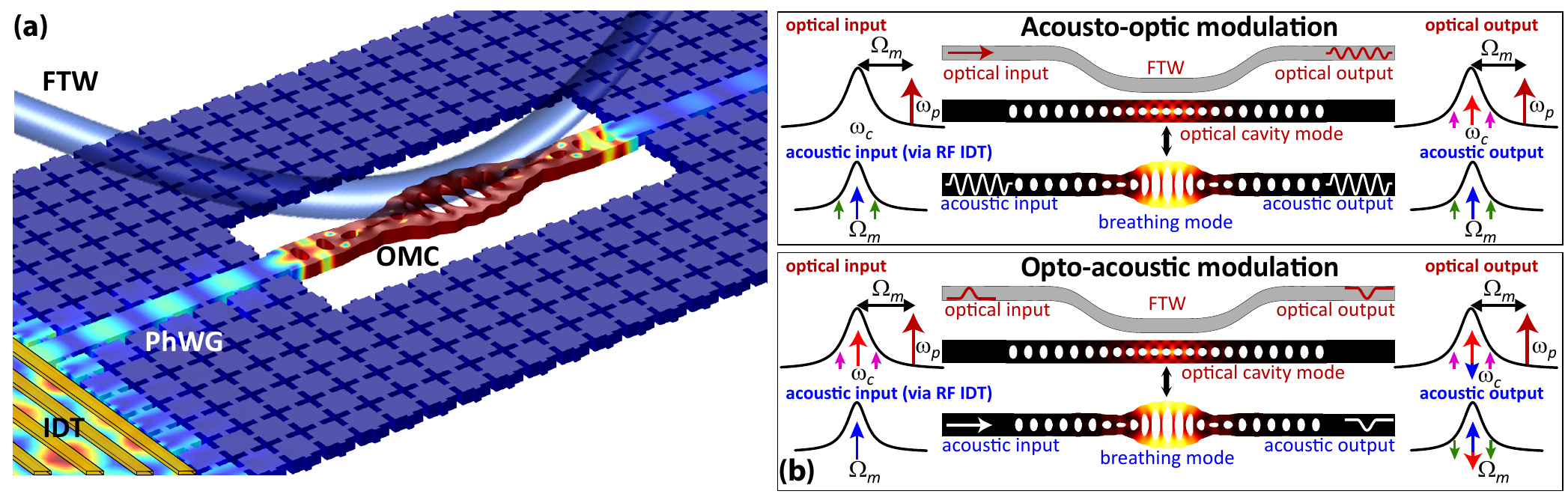}
\caption{(a) Illustration of the piezo-optomechanical circuit platform. A nanobeam optomechanical crystal cavity (OMC) is optically probed through an optical fiber taper waveguide (FTW) and acoustically coupled via a phononic crystal waveguide (PhWG), which in turn is sourced by a radio frequency (RF) interdigitated transducer (IDT). (b) Schematic depiction of two time-domain processes studied in this work, acousto-optic modulation (top) and opto-acoustic modulation (bottom).  In acousto-optic modulation (top), an RF waveform is mixed with an RF carrier at $\Omega_{m}$ and applied to the IDT, while a continuous wave (CW) optical field at $\omega_{p}$, detuned from the optical cavity at $\omega_{c}$ by $\Omega_{m}$, is coupled into the FTW. Due to the photoelastic interaction between the displayed 1550~nm optical mode and 2.4~GHz mechanical breathing mode in the nanobeam OMC, the input RF waveform is mapped onto the output optical field.  In opto-acoustic modulation (bottom), an RF carrier at $\Omega_{m}$ is applied to the IDT while a phase-modulated optical pulse (pump at $\omega_{p}$ and sideband centered at $\omega_{c}$=$\omega_{p}$+$\Omega_{m}$) is coupled into the FTW.  If the phase between the optical pulse and RF carrier is set appropriately, this results in destructive interference in the acoustic domain, with both the output acoustic wave and optical wave bearing the signature of this interaction.  The input and output ports of the FTW and nanobeam OMC indicate the optical and acoustic waveforms in the two different cases, while the relevant sidebands in the optical and acoustic domains are shown on the sides of the central images.}
\label{fig:Fig1}
\end{figure*}

\section{Experimental platform}
Our device architecture~\cite{balram2016coherent} is depicted schematically in Figure~\ref{fig:Fig1}(a).  At its heart is a GaAs nanobeam optomechanical crystal cavity that supports both a localized optical mode ($\lambda$ $\approx$ 1550 nm) and mechanical mode ($f_{m}$ $\approx$ 2.4 GHz), with the simulated modes displayed in Fig.~\ref{fig:Fig1}(b). The optomechanical interaction is engineered using the photoelastic effect~\cite{chan2012optimized,balram2014moving} and the coupling strength ($g_{0}/2\pi$ $\approx$ $1.1$~MHz~$\pm$~60~kHz) achieved in these devices is among the highest measured for an integrated cavity optomechanical system. The mechanical motion of this breathing mode is driven due to three different sources: (i) incoherent Brownian motion due to contact with a finite temperature thermal bath; (ii) coherent RF-driven motion~\cite{bochmann2013nanomechanical,fong2014microwave,balram2016coherent}, in which an RF signal applied to an interdigitated transducer (IDT) excites a propagating surface acoustic wave that travels through a phononic crystal waveguide and preferentially excites the breathing mode when incident on the cavity; and (iii) coherent optically-driven motion~\cite{aspelmeyer2014cavity}, wherein a phase-modulated optical signal ($f_{mod}$ $\approx$ $f_{m}$) drives the mechanical motion (the interference between the optical carrier and sideband produces a beat-note at the mechanical frequency, which drives the cavity through electrostriction). Here, we study this system in the time-domain, focusing on modulation phenomena in both the optical and acoustic domains, mediated by the optomechanical interaction and coupling to both photonic and phononic waveguides.

Our experimental setup is shown in Fig.~\ref{fig:Fig2}, along with an electron microscope image of one of our devices.  An arbitrary waveform generator (AWG) or digital delay generator is used to create various waveforms that can be mixed onto an RF carrier produced by an RF signal generator.  This RF signal can then either be amplified and sent into an IDT, passed through a phase shifter and into an electro-optic phase modulator to generate sidebands on an optical signal that is coupled into the optomechanical cavity using a fiber taper waveguide, or both, depending on the configuration of the switches (labeled 1-4 in the diagram).  The transmitted optical signal past the optomechanical cavity is photodetected with an avalanche photodiode (APD), mixed down to an intermediate frequency by using an IQ demodulator with the RF signal generator output as a local oscillator, amplified, and sent into a 8~GHz bandwidth real-time oscilloscope with a sampling rate of 25~GHz.  The table at the bottom of Fig.~\ref{fig:Fig2} indicates the switch configurations for the different experiments we proceed to describe in the following sections.

\section{Acousto-optic modulation}
\begin{figure}
\includegraphics[width=\linewidth]{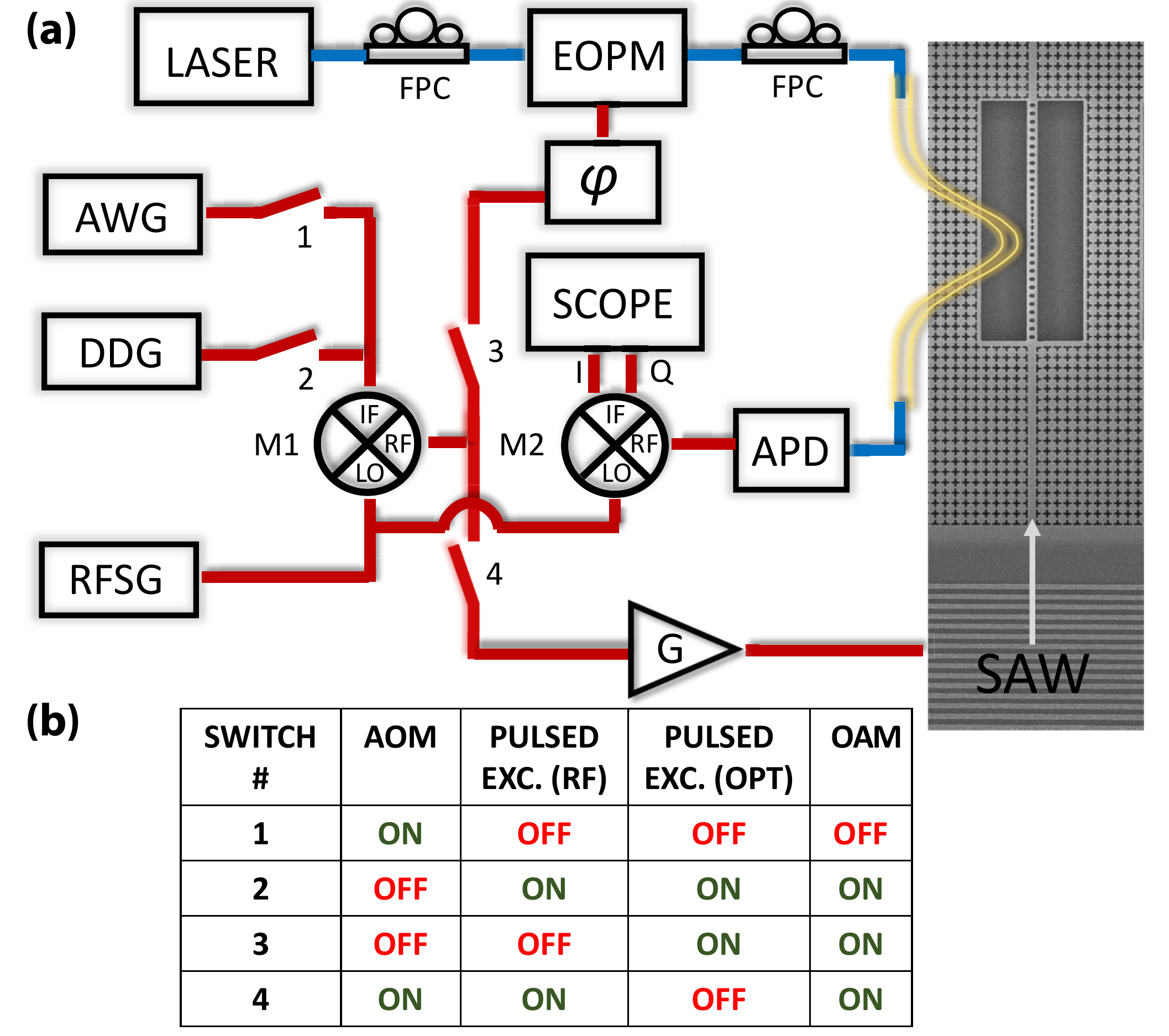}
\caption{(a) Schematic of the experimental setup used for the measurements. AWG: Arbitrary Waveform Generator, DDG: Digital Delay Generator, RFSG: Radio Frequency Signal Generator, EOPM: Electro-Optic Phase Modulator, FPC: Fiber Polarization Controller, SCOPE: High speed digital phosphor oscilloscope, APD: Avalanche Photo Diode, G: RF amplifier, $\varphi$: Phase Shifter, {M1,M2}: RF Mixers, SAW: Surface Acoustic Wave, IF = intermediate frequency, LO = local oscillator. (b) Table illustrating the switch configurations for the different experimental scenarios. AOM: Acousto-optic modulation; Pulsed Exc. (RF): Response of the cavity to RF pulses applied to the IDT; Pulsed Exc. (OPT): Response of the cavity to RF pulses applied to the EOPM; OAM: Opto-acoustic modulation.}
\label{fig:Fig2}
\end{figure}

The ability to coherently drive mechanical motion of the cavity using an applied RF signal allows us to operate the device as a resonant (in both the optical and mechanical domains) acousto-optic modulator (AOM), as depicted schematically in the top panel of Fig.~\ref{fig:Fig1}(b). To understand the device operation, we note that the displacement amplitude of a SAW is determined by the applied RF voltage, and thus modulations in the applied voltage get mapped into modulations of the SAW amplitude. By using such propagating acoustic waves to excite the localized breathing mode of the cavity, we can transfer the RF voltage modulation to cavity mechanical mode displacement modulation. Since the optomechanical cavity converts displacement fluctuations into optical intensity or phase fluctuations (depending on the laser detuning), the transmitted optical signal thus gets amplitude or phase modulated.  In our experiments, we adjust the laser detuning such that the transmitted optical signal is amplitude modulated.

\begin{figure}
\includegraphics[width=\linewidth]{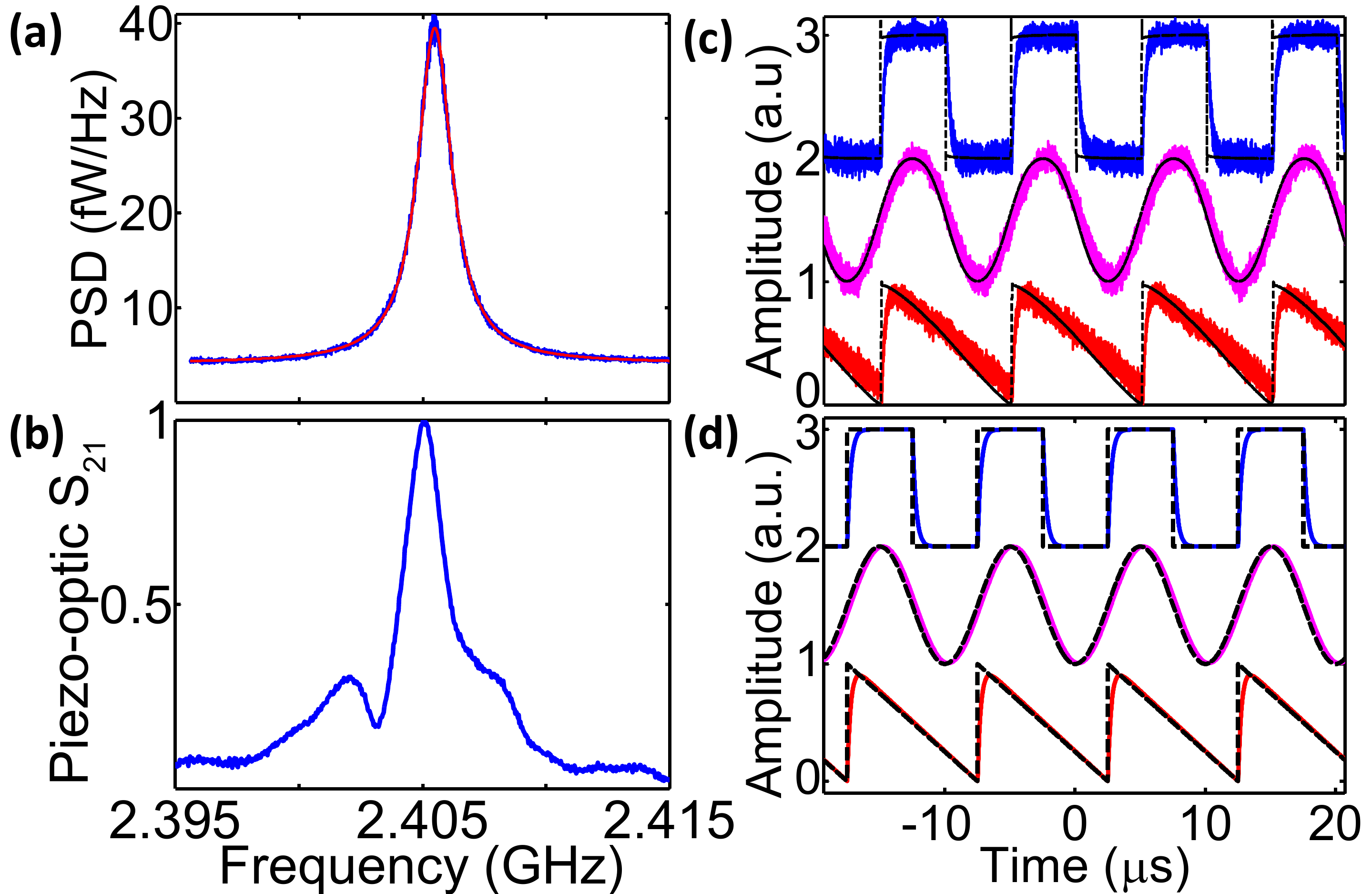}
\caption{(a) Thermal noise spectrum (blue) with Lorentzian fit (red) of the localized breathing mode. (b) Piezo-optic $S_{21}$ spectrum. The resonant peak in (b) corresponds to the peak in (a) showing coherent excitation of the localized mode. (c) Acousto-optic modulation: Square (blue), Sine (magneta), and Sawtooth (red) modulation are imposed on the optical field using a modulated SAW to drive the mechanical motion (the dashed black lines show the applied modulation patterns for reference). (d) Sideband transmission amplitude for the modulation patterns applied in (c), estimated by solving the coupled equations of cavity optomechanics.}
\label{fig:Fig3}
\end{figure}

To operate the piezo optomechanical circuit as an AOM, we use the setup depicted in Fig.~\ref{fig:Fig2} - closing switches 1 and 4 and leaving switches 2 and 3 open.  The AWG produces an intermediate frequency waveform ($IF$) that is mixed with the 2.4~GHz RF carrier ($LO$,) amplified, and applied to the IDT. The photodetected signal is demodulated at the $LO$ frequency using an IQ demodulator, amplified, and sent to two channels of the oscilloscope, whose acquisition is triggered by the AWG.

We first identify the appropriate $LO$ and available bandwidth for $IF$ through thermomechanical and piezo-optical spectroscopy~\cite{bochmann2013nanomechanical,fong2014microwave,balram2016coherent}.  The thermal noise spectrum of our optomechanical cavity, under weak enough optical excitation to avoid any dynamic back-action effects, is shown in Fig.~\ref{fig:Fig3}(a), while the photodetected signal due to coherent motion driven by the RF IDT and through a phononic waveguide, measured using the approach in Ref.~\onlinecite{balram2016coherent}, is shown in Fig.~\ref{fig:Fig3}(b).  We confirm that $LO\approx$~2.4~GHz is needed to drive the IDT to generate a propagating acoustic wave to match the localized breathing mode frequency, and that the breathing mode quality factor limits the bandwidth of $IF$ to $\approx$~1.7~MHz.

As Fig.~\ref{fig:Fig3}(c) demonstrates, the application of different modulation signals in the RF domain (black dashed lines in the figure) results in the corresponding modulations in the photodetected optical signal.  We have shown examples where $IF$ is a square wave (blue), sine wave (magneta), and sawtooth wave (red), and observe faithful signal transduction from the RF to the optical domain, subject to the bandwidth limitations imposed by the mechanical cavity.  The modulation effects can be understood well (Fig.~\ref{fig:Fig3}(d)) by numerically solving the coupled equations of cavity optomechanics, which are discussed in detail in the next section.  We note that in contrast to prior demonstrations of amplitude modulation in which the optical cavity was non-resonantly shaken by an incident SAW~\cite{tadesse2014sub,li2015nanophotonic,de2005modulation}, here we use the phononic waveguide to preferentially excite the localized breathing mode of the cavity, which has a high mechanical quality factor ($Q_{m}$). Moreover, the strong optomechanical coupling rate ($g_0/2\pi\approx1.1$~MHz) allows us to convert efficiently between the displacement fluctuations of the cavity breathing mode and the intensity fluctuations of the transmitted optical signal.  This overall process has the potential to result in acousto-optic modulators that operate at GHz carrier frequencies with low operating voltages.

To quantify the operation of our resonant AOM, we characterize its $V_{\pi}$ (the voltage required to get a $\pi$ phase shift). By taking a ratio of the modulations obtained using a phase modulator with known $V_{\pi}$ and the RF signal applied to the IDT (comparing peak heights of the two CW signals in the power spectrum), we estimate $V_{\pi}$~=~720~mV~$\pm$~72~mV (the error arises primarily from the uncertainty in the reference signal and is a one standard deviation value, as discussed in the Appendix). This estimated value takes into account the $S_{11}$ (reflection) spectrum of the IDT. By optimizing the acoustic energy transfer between the SAW and the nanobeam breathing mode and improving the mechanical quality factor $Q_{m}$, we can push $V_{\pi}$ below 1 mV. Such low $V_{\pi}$ AOMs can potentially be useful for mapping weak electromagnetic signals from the RF to the optical domain~\cite{bagci2014optical}, where they can propagate over long distances in optical fiber and be sensed by single-photon detectors.  This functionality may have use in areas like radio astronomy~\cite{rohlfs2013tools}, MRI~\cite{kovacs2005cryogenically}, and radar~\cite{eaves2012principles}.

\section{Optical drive, acoustic wave drive, and dynamical back action}

\begin{figure*}
\includegraphics[width=0.8\linewidth]{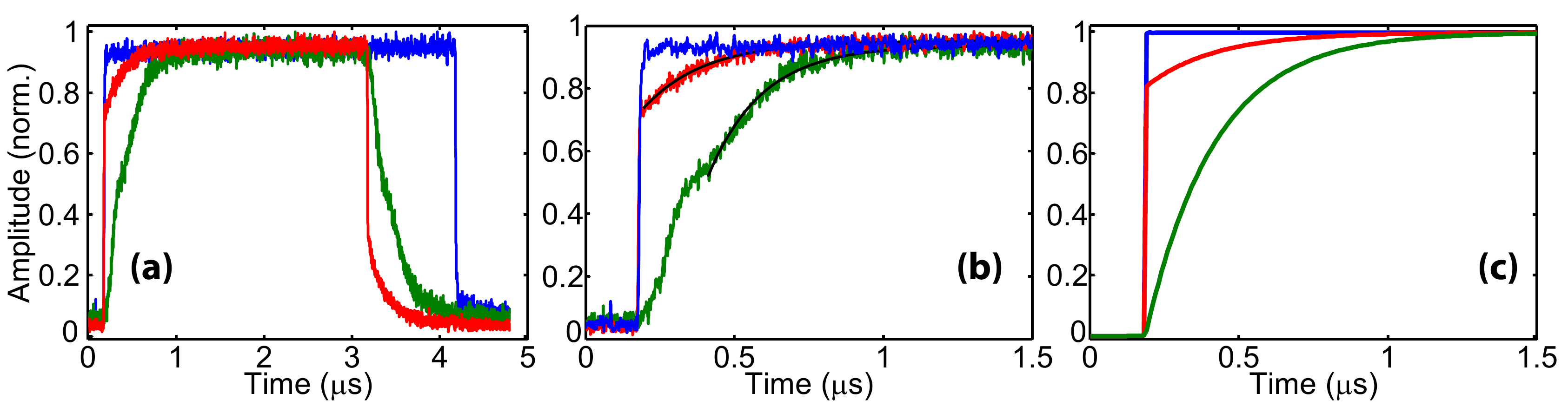}
\caption{(a) Measured system response to pulsed excitation. Blue: Baseline case in which the system is driven optically and off mechanical resonance; Red: On-resonance optically-driven motion; and Green: RF-driven motion due to the IDT-generated propagating acoustic wave. In comparison to the off-resonance optically-driven case, on-resonance optically-driven motion has a slower rise time due to the mechanical cavity ring-up. (b) Zoom-in of (a) showing the rising edges more clearly. The solid lines in (b) are nonlinear least squares fits to the data assuming a single exponential decay time. (c) Predicted system response to pulsed off-resonance optical excitation (blue), pulsed on-resonance optical excitation (red), and pulsed RF excitation (green), determined by solving the coupled (optical and mechanical) equations of cavity optomechanics.}
\label{fig:Fig4}
\end{figure*}

To further explore the dynamics of our cavity optomechanical system, we excite the localized breathing mode of the cavity using RF and optical pulses and optically monitor the cavity displacement. The experimental setup is similar to that used for the AOM experiments in the previous section, except that we replace the arbitrary waveform generator with a pulse generator. The mechanical cavity displacement is probed by measuring the modulation induced on the transmitted optical signal. The mechanical cavity can be driven either through the RF channel (green curve in Fig.~\ref{fig:Fig4}(a)) using an IDT (Fig.~\ref{fig:Fig2}; switches 2 and 4 closed, and 1 and 3 open) or through the optical channel (blue and red curves in Fig.~\ref{fig:Fig4}(a)) by phase modulating the optical input (Fig.~\ref{fig:Fig2}; switches 2 and 3 closed, and 1 and 4 open), where the interference between the optical input and the phase modulated sideband drives the mechanical motion using electrostriction).

To understand the response for the different driving conditions, we can start with the optical drive. The blue curve in Fig.~\ref{fig:Fig4}(a) corresponds to the case when the RF carrier signal ($LO$) is set to a frequency that is detuned off the mechanical resonance. In this regime, the system effectively has an infinite mechanical bandwidth (no $Q_{m}$) and the transmitted cavity signal shows a sharp rising edge. This situation corresponds exactly to putting a phase modulated signal on a bare optical cavity and measuring the transmitted amplitude modulated (AM) response (which has a rise time corresponding to the optical cavity lifetime). On the other hand, when the cavity is driven at its mechanical frequency (red curve in Fig.~\ref{fig:Fig4}(a)), we can clearly see an evidence of dynamic back action. The probe transmission now has two time scales; a sharp step corresponding to the AM induced on the directly transmitted optical signal (this is similar to what happens in the off-resonance case) and a second slower rise that occurs due to coherent scattering of the optical pump as a result of cavity motion. To put this result in context, we note that in the frequency domain, this corresponds to the observation of coherent dynamic back action leading to optomechanically-induced transparency~\cite{weis2010optomechanically,safavi2011electromagnetically}.

When the system is driven through the RF channel by an IDT (green curve in Fig.~\ref{fig:Fig4}(a)), we see a slower response. As opposed to the optical driving case where the input optical signal was phase modulated, the input optical signal here is continuous wave and the phase modulation is induced by mechanical motion of the cavity that is driven by the phononic waveguide.  The timescale is thus set by the loaded mechanical quality factor of the cavity ($Q_{m}=1428\pm2$, where the uncertainty is given by the 95~$\%$ confidence interval of the Lorentzian fit shown in Fig.~\ref{fig:Fig3}(a)). The time constants extracted from fitting the curves in Fig.~\ref{fig:Fig4}(b) are $183$~ns~$\pm$~3~ns for the RF excitation and $205$~ns~$\pm$~4~ns for the optically driven excitation (model fit uncertainty, 95~$\%$ confidence interval). These numbers agree reasonably well with the time constant ($190$~ns~$\pm$~1~ns) extracted from the aforementioned value for $Q_{m}$.

We can understand and model the observed effects by using the coupled equations of cavity optomechanics:

\begin{eqnarray}
\dot{a} = -(i\Delta+{\frac{\kappa_{i}}{2}})a-ig_{0}a(b+b^{\dagger})-\sqrt{\frac{\kappa_{e}}{2}}a_{in}
\label{eq:intracavity_opt_field}
\end{eqnarray}

\begin{eqnarray}
\dot{b} = -(i\Omega_{\text{m}}+{\frac{\gamma_{i}}{2}})b-ig_{0}a^{\dagger}a-\sqrt{\frac{\gamma_{e}}{2}}b_{in}
\label{eq:intracavity_mech_disp}
\end{eqnarray}

\noindent with $a(a^{\dagger})$ representing the annihilation (creation) operator for the intracavity optical field and $b(b^{\dagger})$ the annihilation (creation) operator for the mechanical displacement. $\Delta$ represents the detuning of the control beam from the optical cavity resonance frequency, $\kappa_{i}$ the intrinsic decay rate of the optical cavity, and $\kappa_{e}$ is the extrinsic decay rate (coupling rate) to the waveguide. $\Omega_{m}$ represents the mechanical mode frequency, $\gamma_{i}$ the intrinsic decay rate of the mechanical cavity, and $\gamma_{e}$ is the extrinsic decay rate (coupling rate) to the phononic waveguide. $a_{in}$ and $b_{in}$ represent the optical and acoustic field drive strengths in the photonic and phononic waveguides, respectively, while $g_{0}$ is the vacuum optomechanical coupling rate.

We assume that we are operating in a linearized regime of small perturbations around some steady state. This allows us to model the dynamics by Fourier transforming the input signal, finding the system response to each frequency component (with appropriate phase) and inverse Fourier transforming the output. Figure~\ref{fig:Fig4}(c) shows the results of our calculations, which faithfully reproduce the response for the optical driving (both on and off resonance) and produce the right time scales for the mechanical response.  Figure~\ref{fig:Fig4}(b) shows the experimental data plotted over the same range of timescales, indicating the correspondence with the simulations.  This correspondence is also observed in the RF-driven AOM data and simulations from Fig.~\ref{fig:Fig3}(c)-(d).

While the optically-driven response matches theory well, a closer inspection of the zoomed-in RF-driven data in Fig.~\ref{fig:Fig4}(b) reveals some deviation from the picture we have described.  In particular, the RF-driven response is not perfectly described by a single timescale, but instead seems to include a fast component and then a slower component (we note that this fast component is still much slower than the fast component in the optically-driven case). We believe the fast rise originates due to imperfect excitation of the mechanical breathing mode.  While the localized breathing mode is most strongly excited by the incident acoustic wave, it can be accompanied by excitation of a variety of beam modes at the same frequency (which have low $Q_{m}$), and these modes can also modulate the optical cavity and give rise to a fast component. In effect, in the fast regime, the system behaves like a photonic cavity that is non-resonantly modulated by the acoustic wave, which is essentially the mechanism demonstrated in previous acousto-optic modulators based on manipulating a photonic cavity with surface acoustic waves~\cite{de2005modulation,tadesse2014sub,li2015nanophotonic}.  As our model neglects the potential excitation of low-$Q_{m}$ modes, it fails to capture the resulting fast rise seen in the experiments.  On the other hand, when the cavity is driven optically, only the localized breathing mode is excited because it is the only mechanical mode that has strong $g_{0}$, and our model fully reproduces the experimental results.

\section{Opto-Acoustic Modulation}

Thus far, we have probed the dynamics of our cavity optomechanical system under pulsed excitation through either the optical channel or the RF channel. However, as discussed in steady-state measurements in Ref.~\onlinecite{balram2016coherent}, these two excitation processes can lead to interference effects in the acoustic domain as long as phase coherence is maintained between the two drive channels. In particular, by choosing the amplitude and phase of the RF excitation appropriately, one can observe destructive interference between the optically-driven and RF-driven motion.  Such interference, schematically depicted in Fig.~\ref{fig:Fig1}(b), when taken together with the ability to route acoustic waves into and out of the optomechanical cavity through phononic waveguides, results in opto-acoustic modulation - namely, the ability to gate the propagation of acoustic waves through the application of an optical control field. Although the gating requires phase coherence between the RF and optical channels, the ability to control acoustic wave propagation with optical fields is of significant interest considering the difficulty of dynamically manipulating acoustic waves through other means.  In particular, while in photonics, the refractive index can be readily manipulated through free carriers, thermo-optic dispersion, and nonlinear optical processes (e.g., the Kerr effect), manipulation of acoustic wave signals would require manipulation of a material's density or elastic constants, which are in general considerably less sensitive to external perturbations.

\begin{figure}
\includegraphics[width=\linewidth]{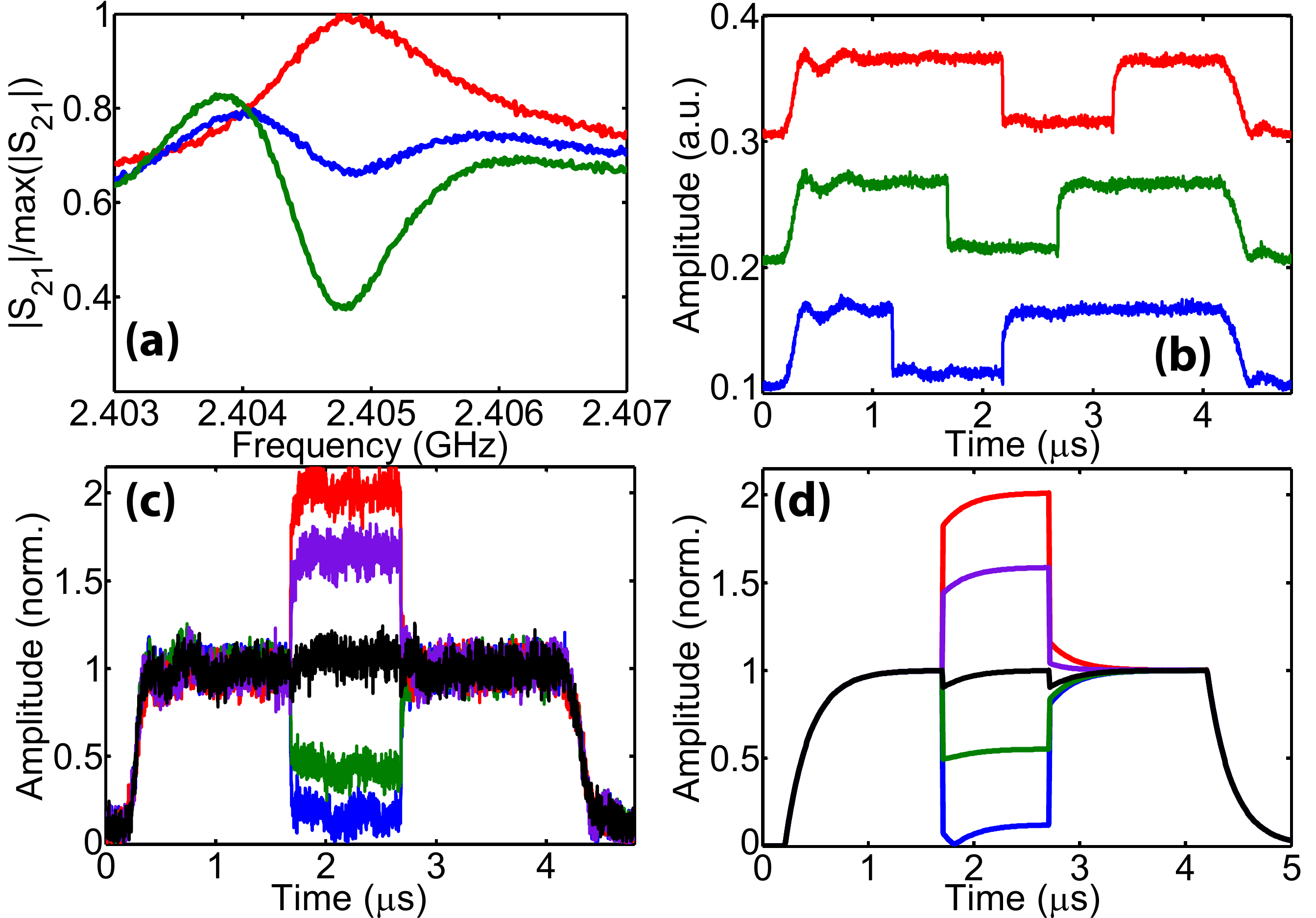}
\caption{(a) Normalized piezo-optic $S_{21}$ response of the cavity for three scenarios (see text for detailed explanation). Red: Optomechanically Induced Transparency (OMIT), blue: Acoustic Wave Interference, green: Electromechanically Induced Transparency (EMIT). (b) Pulse position modulation (gated interference) of 4 ${\mu}s$ long acoustic pulses using 1 ${\mu}s$ optical pulses with varying delay between the pulses (a vertical offset between the traces is used for clarity). (c) Mechanical cavity response to simultaneous excitation by acoustic pulses (4~$\mu$s length) and optical pulses (1~$\mu$s length) with varying RF phase $\varphi$ between them. The central case shown in (b) corresponds to the blue trace. (d) Predicted system response to the pulses in (c).}
\label{fig:Fig5}
\end{figure}

Figure~\ref{fig:Fig5}(a) plots the transmitted probe sideband amplitude, which is effectively an optical readout of the mechanical cavity displacement amplitude, in the continuous wave (both in the RF and optical domains) case. Starting with no RF-driven motion (red curve in Fig.~\ref{fig:Fig5}(a)), we observe optomechanically induced transparency (OMIT)~\cite{weis2010optomechanically,safavi2011electromagnetically} when the system is driven by a phase-modulated optical signal. The transparency in the probe (sideband) transmission occurs due to the interference between the directly transmitted probe signal and the coherently scattered signal from the pump due to the mechanical motion induced in the cavity. By turning the RF power on and setting the phase for destructive interference, we observe a cancellation of the optomechanically-induced transparency (blue curve in Fig.~\ref{fig:Fig5}(a)) on mechanical resonance, because the coherent motion of the cavity is zero and the scattering from the carrier to the sideband gets suppressed. Further increasing the RF power converts the transparency to an absorption dip (green curve in Fig.~\ref{fig:Fig5}(a), corresponding to electro-mechanically induced transparency~\cite{bochmann2013nanomechanical,fong2014microwave}) as the scattered signal is now $\pi$ out of phase. We can use this interference effect to demonstrate optical gating of acoustic pulses.

Figure~\ref{fig:Fig5}(b) shows the results of such experiments. We use 1~${\mu}$s phase-modulated optical pulses to gate the propagation of 4~${\mu}$s long acoustic pulses. The same RF carrier signal is fed into the two channels with a phase delay (set for destructive interference), as depicted in Fig.~\ref{fig:Fig2}, where switches 2, 3, and 4 are closed and switch 1 is left open. By varying the time delay between the two pulses, one can carve out arbitrary 1~${\mu}$s chunks of the acoustic pulse using the optical pulse, as shown in Fig.~\ref{fig:Fig5}(b). We note that the plotted amplitude in Figure~\ref{fig:Fig5}(b)-(d) is an optical readout of the intracavity phonon population, but the same behavior will hold for the transmitted phonons as well (though any directly transmitted acoustic wave component, uncoupled to the mechanical cavity, will reduce the contrast). While the results in Fig.~\ref{fig:Fig5}(b) represent the interaction for a phase difference ($\pi$) fixed for destructive acoustic wave interference, we can map out the response for the other phases as well, ranging from constructive interference (red curve in Fig.~\ref{fig:Fig5}(c)) to destructive interference (blue curve in Fig.~\ref{fig:Fig5}(c)), with intermediate cases shown as well. The experiment can be modelled well using the method discussed in the previous section, and the corresponding results are shown in Fig.~\ref{fig:Fig5}(d).

\section{Conclusions}
We have characterized the dynamic response of GaAs piezo-optomechanical circuits to pulsed excitation in both the RF and the optical domain, where the former drives the central optomechanical cavity through an in-coupling acoustic waveguide, while the latter uses electrostrictive forces. Driving the cavity through the RF channel allows us to operate the device as a doubly resonant (for both the optical and mechanical modes) acousto-optic modulator with a $V_{\pi}=720$~mV $\pm$ 72~mV. Moreover, the pulsed response allows us to visualize the dynamic back action and account for the difference in the excitation characteristics through the two channels. We also use acoustic wave interference between the RF and optically driven motion to interferometrically gate the propagation of acoustic pulses through the system using optical pulses.

While RF signals have been used to excite mechanical motion in this work, our measurements focus on optical detection of the intra-cavity motion. This utilizes the exquisite sensitivity of cavity optomechanical systems to minute perturbations. In the future, we would like to extend this to detecting the transmitted / propagating phonon component, which can be done electrically through IDTs provided that the intracavity motion is efficiently transferred to an acoustic wave whose propagation characteristics are suitable for exciting the IDT.  This requires further engineering of the coupling region between the propagating and localized acoustic wave resonances to reach the overcoupled regime~\cite{fang2016optical}, and improving the overlap between the propagating acoustic wave and the IDT, for example, through the use of curved IDT geometries~\cite{wu2005analysis}.  Such efforts would be key to demonstrating signal processing functionalities in which optical fields are used to manipulate radio frequency waves through acoustic tranducers, and more specifically, to efficient and bi-directional microwave-to-optical conversion.

\section{Acknowledgments}

K.C.B. acknowledges support under the Cooperative Research Agreement between the University of Maryland and NIST-CNST Award 70NANB10H193. J.D.S. acknowledges support from the KIST flagship institutional program.

\textbf{\center{\appendix{Appendix A: Acousto-Optic Modulator $V_{\pi}$}}}

\hspace{0.1in}

\begin{figure}
\includegraphics[width=\linewidth]{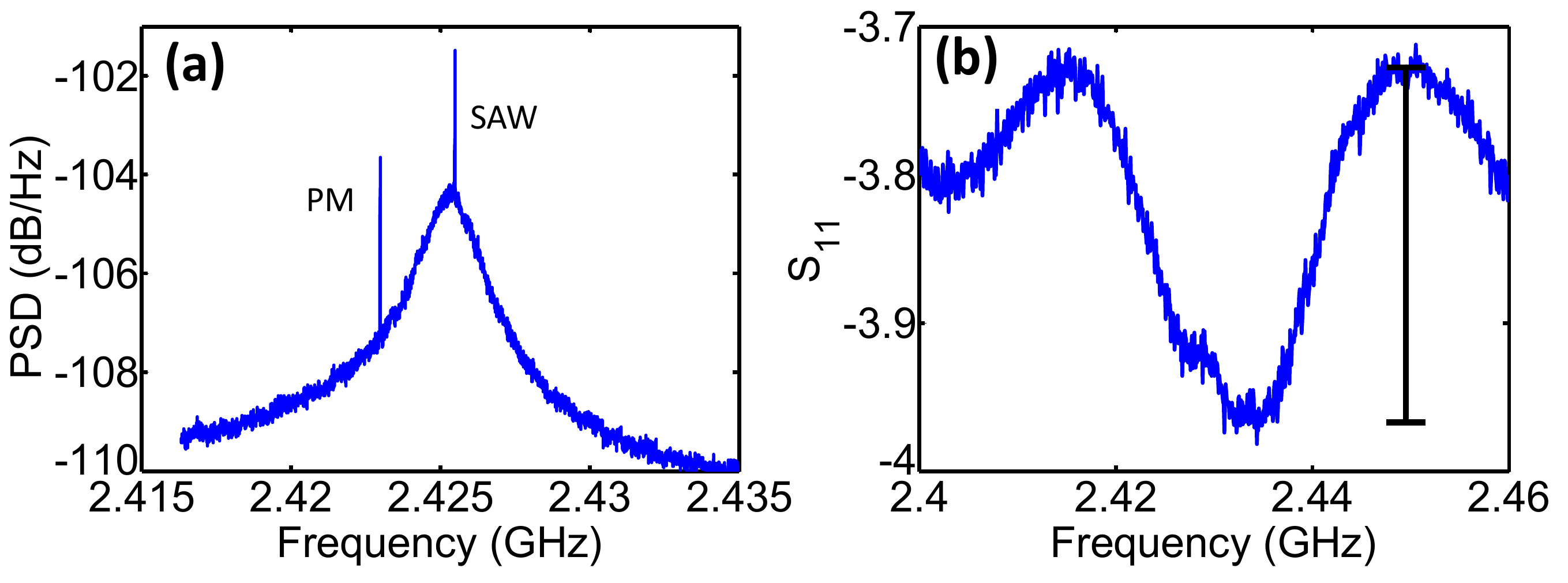}
\caption{(a) Mechanical mode thermal noise spectrum with SAW and phase modulator (PM) calibration tones indicated. By comparing the peak heights, we can estimate the modulation index imposed by the SAW. 0~dB is referenced to 1~mW of power (i.e., in dBm). (b) Electrical reflection spectrum of the IDT, with the resonance contrast used to estimate the voltage that is coupled to the IDT and generates a propagating SAW.}
\label{fig:FigS1}
\end{figure}

To determine the equivalent $V_{\pi}$ of our piezo-optomechanical circuits when operating them as acousto-optic phase modulators, we use the relationship:

\begin{eqnarray}
V_{\pi} = \pi\frac{V_{\text{SAW}}}{\beta_{\text{SAW}}}
\end{eqnarray}

\noindent where $V_{\text{SAW}}$ is the applied RF voltage to the IDT and $\beta_{\text{SAW}}$ is the modulation index imposed by the propagating SAW generated by the IDT.  To determine $V_{\text{SAW}}$, we use the resonant dip in the $S_{11}$ spectrum of our IDT to determine the transmitted RF voltage (in our case, the contrast is 0.25 dB, as seen in Fig.~\ref{fig:FigS1}(b)) and a 100~$\Omega$ load (the IDT impedance can be extracted from the $S_{11}$ spectrum) to determine the corresponding voltage. By taking the difference between on- and off-resonance reflection, we remove the background reflection that occurs due to an electrical impedance mismatch between a 50~$\Omega$ transmission line and the IDT.  In other words, our estimate of $V_{\pi}$ is valid for a scenario in which the IDT is impedance-matched to the RF excitation line.  Achieving this in practice requires the use of balun elements for impedance matching.

To determine $\beta_{\text{SAW}}$, we compare the peak height induced by the SAW modulation to that induced by a known optical modulation (applied using an EOPM) in the photodetected spectrum. An example spectrum is shown in Fig.~\ref{fig:FigS1}(a). Alternately, we can also determine $\beta_{\text{SAW}}$ by comparing the height of the SAW-induced peak in the photodetected spectrum to the area of the thermal noise spectrum (broad Lorentzian curve in Fig.~\ref{fig:FigS1}(a)), with the thermal noise spectrum being generated by contact of the mechanical resonator with the room temperature environment at 298~K (the sample is tested in ambient conditions). The error in the $V_{\pi}$ of the acousto-optic modulator ($\sigma V_{\pi}$) is dominated by the error in the $V_{\pi}$ of the reference EOPM ($\sigma V_{\text{sig,PM}}$).

\begin{eqnarray}
V_{\pi} = \pi\frac{V_{\text{SAW}}}{C\beta_{\text{PM}}}
\end{eqnarray}

\begin{eqnarray}
V_{\pi} = \pi\frac{V_{\text{SAW}}V_{\pi,\text{PM}}}{C{\pi}V_{\text{sig,PM}}}
\end{eqnarray}
where $\beta_{\text{PM}}$ is the modulation index of the phase modulator, $C$ is a constant representing the ratio of the peak heights, and $V_{\text{sig,PM}}$ is the applied RF voltage to the PM. Hence:

\begin{eqnarray}
({\sigma}V_{\pi,\text{AOM}}) = \pi\frac{V_{\text{SAW}}{(\sigma}V_{\pi,\text{PM}})}{C{\pi}V_{\text{sig,PM}}}
\end{eqnarray}

%

\end{document}